# Dissipative structure formation during crystallization of alloys under high-nonequilibrium conditions

L. P. Tarabaev[1] and V. O. Esin[1]

[1]*Institute of Metal Physics, Ural Division of the Russian Academy of Sciences, 18, S. Kovalevskaya St. GSP-170, Ekaterinburg, 620041, Russia*
*E-mail: tarabaev@imp.uran.ru, yesin@imp.uran.ru*

## Abstract

The crystallization of a binary system is investigated in computer model which takes into account a temperature dependence of diffusion coefficient and a nonequilibrium partition of dissolved component of the alloy. The dependence of interface velocity V on an undercooling ΔT at the dendrite tip is obtained during rapid solidification of Fe-B and Ni-B systems. The morphological transition which is conditioned by change of a diffusion growth regime on thermal growth at some undercooling ΔT ≥ ΔT* is detected. This transition is characterized by discontinuous magnification of the dendrite growth velocity. Values of a critical undercooling ΔT* and a growth velocity discontinuity depend both on a degree of an anisotropy of a kinetic coefficient, and on difference in energies of activation for atomic kinetics and for diffusion on an interface.

## 1 Introduction

The process of crystallization in a system far from equilibrium has features, which manifest themselves in the morphology, crystal growth velocity, and segregation of dissolved alloy components. So under conditions of high cooling rates of melt (R ~ $10^6$ K/c), when the deep undercoolings are reached an irregular morphology of solidification, nonequilibrium «trapping» of impurity, and neighborhood of crystalline and amorphous phases are observed [1]. For sufficiently high growth velocities, i.e. for certain critical undercooling the sharp transition to a partitionless regime of crystallization will take place [2]. It afterwards was called as kinetic phase transition [3-5]. The critical undercooling can reach large values: so for Ni-B alloy of ~ 200 ÷ 300 K [6]. The dissipative structures formed in such system, essentially influence on set of main properties of prepared material.

The method of computer simulation [7] allows to study the formation of a complex morphology of the solid – liquid interface and it dynamics during a crystallization of pure metals [8] and metal alloys [9, 10]. In this work the crystallization from one centre of a binary essentially nonequilibrium system is investigated in computer model [11] which takes into account the temperature dependence of the diffusion coefficient and the nonequilibrium partition of dissolved component of the alloy [12].

## 2    Computer model

The computer model is based on a finite difference method. The two–dimensional finite-difference grid divides the system into cells. Each cell is characterized by a volume fraction of a solid phase $g_S$. Assuming the normal mechanism of crystal growth the velocity of an interface motion $V$ in a two-phase cell ($0 < g_S < 1$) can be written as follows:

$$V = \beta \Delta T, \qquad (1)$$

where β is anisotropic kinetic coefficient, $\Delta T = T_E - T_I = T_M(1 - d_0\kappa) - mc_I - T_I$ is kinetic undercooling at the interface; $T_E$ is equilibrium temperature, $T_M$ is temperature of melting of first component, $d_0 = \gamma_{SL}/Q$ is capillary length, $\gamma_{SL}$ is surface tension of crystal – melt interface, $Q$ is heat of melting, κ is interface curvature, $m$ is the slope of the equilibrium liquidus line (without sign), $T_I$ and $c_I$ are the temperature and the concentration in the liquid at the interface, respectively.

The nonequilibrium effect of solute partition at interface is described by expression [12] for partition coefficient $k$:

$$k(V) = \frac{c_S}{c_L} = \frac{V/V_D + k_e}{V/V_D + 1 - (1 - k_e)c_L}, \qquad (2)$$

where $c_S$ and $c_L$ are concentration in the solid and in the liquid at the interface, respectively, $k_e = k_0/k_0^A$; $k_0$ and $k_0^A$ are the equilibrium partition coefficients of solute and solvent, respectively, $V_D \equiv fva\exp(-E_a/RT) = fD/a$ is the rate of diffusion; $f$ is geometric factor, $v$ is the atomic vibration frequency, $a$ is interatomic spacing, $E_a$ is the activation barrier for diffusion through the interface, $D$ is coefficient of diffusion at the interface. Rate of diffusion is the ratio of the diffusion coefficient at interface to the interatomic spacing. The partition coefficient depends on the ratio of velocity of crystallization $V$ to rate of diffusion $V_D$.

The kinetic effect includes both temperature and orientation dependences of the kinetic coefficient, which polar diagram has the four-fold symmetry, and the directions of maxima coincide with the principal grid directions. Then the velocity of an interface motion $V$ can be written as [1, 4]:

$$V = \beta \Delta T = f'va\exp(-E'_a/RT)Q\Delta T/RT_E T, \qquad (3)$$

where the kinetic coefficient $\beta = f'va\exp(-E'_a/RT)Q/RT_E T$; $f'$ is a factor of anisotropy, $E'_a$ is the activation barrier for atomic kinetics. The ratio of the interface velocity to the diffusion rate in Eq. (2) can be written as follows

$$\frac{V}{V_D} = \frac{f'}{f}\exp\left(-\frac{E'_a - E_a}{RT}\right)\frac{Q\Delta T}{RT_E T}, \qquad (4)$$

and in the case when the energy of activation for atomic kinetics $E'_a$ is equal to the energy of activation for diffusion $E_a$ at the interface from Eq. (4) follows that

$$\frac{V}{V_D} = \frac{f'}{f}\left(\frac{Q\Delta T}{RT_E T}\right). \qquad (5)$$

The ratio $V/V_D$ characterizes the degree of deviation from equilibrium of the interface for a given temperature $T$ ($\Delta T/T$) and entropy of melting ($Q/RT_E$). That is in terms of the atomic kinetics it signifies the ratio of a resulting flux of atoms to an exchange (equilibrium) flux at the interface. And the ratio $f'/f$ characterizes the degree of the anisotropy of growth rate of crystal. In case of $V = V_D$ from the Eq. (5) the expression for a undercooling follows:

$$\Delta T^* = \frac{T_E}{1+(f'/f)(Q/RT_E)}, \qquad (6)$$

which can be the criterion of transition to nonequilibrium trapping of dissolved component of the alloy at interface at undercoolings larger than this value.

*Heat -, mass transfer in a system.* For each volume element $\Omega$ of a system from conservation laws follow the equations for fields of $c_L$, $c_S$:

$$\frac{\partial c_L}{\partial t} = \frac{(1-k)c_L}{g_L}\frac{\partial g_S}{\partial t} + \frac{1}{g_L\Omega}\int_S \left(g_L D_L(T)\vec{\nabla}c_L, d\vec{S}\right), \qquad (7)$$

$$\frac{\partial (c_S g_S)}{\partial t} = kc_L \frac{\partial g_S}{\partial t}, \qquad (8)$$

and $T$:

$$\frac{\partial T}{\partial t} = \alpha \nabla^2 T + \frac{Q}{C}\frac{\partial g_S}{\partial t}, \qquad (9)$$

where the diffusion coefficient in the melt depends on the temperature: $D(T) = a^2 v \exp(-E_D/RT)$; the diffusivity $\alpha$ is accepted identical in both phases; $C$ is heat capacity. The source in (9) (also in (7)) is simulated by algorithm developed in [7], using the expression for the change of a volume fraction of a solid phase in two-phase cell of a system:

$$\frac{\partial g_S}{\partial t} = \frac{V(\vec{n})l(\vec{n})}{\Omega}, \qquad (10)$$

where $\vec{n}$ is the local normal to the interface segment in a two-phase cell and $l(\vec{n})$ is the area of the interface segment. The finite-difference scheme of the problem was formulated with regard for these equations, and the corresponding computer program was modified [11].

Graphic presentation of the equations (1) - (6) for the system Fe-B is shown in Fig.1. We now use the dimensionless quantities: $\tilde{V} = V/v_0$, $\tilde{V}_D = V_D/v_0$, $\Delta \tilde{T} = \Delta T\, C/Q$, where $v_0 = \beta_0 Q/C$ and $\beta_0$ is isotropic kinetic coefficient at the phase equilibrium temperature. We assume that $V_D(T_E)/v_0 = \varepsilon \Theta/(Q/RT_E) = 0.9$, where $\varepsilon$ is the factor which take into account the difference between activation barriers for atomic kinetics and for the diffusion at the interface (ε = 0.31). Dependence of growth velocity $\tilde{V}$ on undercooling of the melt $\Delta\tilde{T}$ is calculated in a maximum (+) for $(f'/f)_{max}$ = 1.5 and in minimum (-) for $f'/f$ = 1 of orientation dependence of a kinetic coefficient, which also depends on the temperature. All these presented above effects are known separately, but in this model they are interdependent. Taking into account the diffusion as the limiting factor, the true curve will lie below than these dependences. The kind of a curve can be obtained as a result of the computer simulation. Crystal growth is controlled by the joint action of kinetic phenomena at the interface and heat transfer and second-component mass transfer in the system.

## 3   Results of computer simulation

The computer simulation of crystallization of the melt from one centre in the system 250 × 250 cells is realized for various initial and boundary conditions: the system is at the given undercooling $\Delta T = \Delta T_{bath}$ with adiabatic boundary and the temperature on system boundaries decreases from some initial value $T_{init} < T_E$ to $T = T_B$ with the given rate of melt cooling $R$. The parameters of the problem approximately correspond to Fe-B system [15, 16]: the iron melting temperature is $T_M^{Fe}$ = 1809 K, the liquidus temperature is $T_E$ = 1803 K at an initial boron concentration $c_0$ = 0.04 wt % B ($\Theta = (C/Q)T_E$ = 2,96); $E_a$ = 83.8 kJ/mol ($E_a/RT_E$ = 5,55), $Q$ = 15.38 kJ/mol ($Q/RT_E \approx 1$), $Q/C$ = 609K, the equilibrium partition coefficient is $k_0$ = 0.015, and the slope of the liquidus line $m$ = -102 K/wt %. $D(T_E)$ = 5 × 10$^{-5}$ cm$^2$/s, $\gamma_{SL}$ = 120 × 10$^{-7}$ J/cm$^2$, α=0.7 × 10$^{-1}$ cm$^2$/s.

Figure 2 shows the morphology of the growing crystal and the temperature field (in relative magnitudes $TC/Q$) in the system in certain moments of the time for various values of the bath undercooling. With increase of the bath undercooling the morphology of growth of crystal is changed from the globular form of diffusion–limited growth (Fig. 2a) to the cellular-dendritic form (Fig. 2b, 2c) and, then, to the needle-like form of thermally controlled growth (Fig. 2d). The change of the morphology from the dendrite with a cellular lateral surface to the needle-like dendrite is shown in Fig. 2c The change in the crystal growth regimes is illustrated by the corresponding changes in the temperature field configuration. The temperature field (Fig. 2c, 2d) indicates that the dendritic growth occurs in a thermally controlled regime. The isotherms are distorted under influence of the crystallization heat releasing. In the melt far from the surface of growing crystal the temperature field is concentric isolines.

To study the dynamic behavior of interface during the evolution of the Fe-B system the change of the crystallization velocity $V$ and undercooling $\Delta T$ at the dendrite tip is calculated. Trajectories of the dendrite tip in the space of variables $V$ and $\Delta T$ are shown in Fig. 3. That is to say this is a phase portrait of a dendrite. Each point corresponds to the state of the interface ($\widetilde{V}, \Delta \widetilde{T}$) in a cell which is on the path of movement of the dendrite tip. The points are obtained by averaging of a undercooling and growth rate over the number of temporary steps, for which the two-phase cell containing the dendrite tip becomes completely solidified. The value of a undercooling $\Delta T$, for which $V = V_D$ in a maximum of kinetic coefficient, is designated as $\Delta T^*$ ($\Delta T^* = 303$ K or $\Delta \widetilde{T}^* = 0.499$). Results of computer simulation obtained both under conditions of melt cooling on the system boundaries until $\Delta T = \Delta T_B$ and under adiabatic boundary conditions of system with initial undercooling $\Delta T = \Delta T_{bath}$ show that the $V$ versus $\Delta T$ curve has an S-like character.

The growth of crystal at undercooling $\Delta T < \Delta T^*$ is limited by the diffusion of the dissolved component rejected by the interface because of a significant separation effect, since the crystallization velocity $V$ is lower than the maximum possible velocity in the kinetic regime ($V < V_D$) and, therefore, the partition coefficient $k(V)$ weakly differs from the equilibrium value, which is much smaller than unity. The velocity jump at a dendrite tip undercooling $\Delta T \geq \Delta T^*$ corresponds to the morphological transition conditioned by the change from a diffusion to a kinetic growth regime controlled by heat transfer in the system. The segregation of the dissolved component at the dendrite tip is practically absent, since the partition coefficient $k(V)$ is close to unity.

Results of computer simulation obtained under melt cooling on the system boundary with the rate R = 0.001 until $\Delta T = \Delta T_B$ with the parameters approximately correspond to Ni-B system $\theta = 3.651$, $E_a/RT_E = 5.55$, $Q/RT_E = 1.2$, $k_e = 0.015$ and $V_D(T_E)/v_0 = 0.9$ and data of experiment for Ni – B [6] are shown in Fig 4. To compare our results of computer modeling to experimental data we used values: $v_0 = 100$ m/s at $\beta_0 = 0.21$ m/s K and $Q/C = 472.65$ K. In the experiment the solidification occurs from multitude of the nucleus therefore crystals because of mutual influence morphologically are not developed, but the data on growth velocity specify the onset of the transition to the partitionless regime. The obtained value of a critical undercooling $\Delta T^* = 244$ K on an order of magnitude will be agreed with the experimental data, in particular, for an alloy Ni - 1 at % B it has been established, that the growth velocity sharply increases and also solidification becomes almost partitionless at critical undercooling $\Delta T^* = 267$ K.

# 4  Conclusions

We have proposed a computer model which takes into account a temperature dependence of the diffusion coefficient and the nonequilibrium partition of dissolved component of the alloy. The dependence of interface velocity V on an undercooling ΔT at the dendrite tip is obtained during rapid solidification of Fe-B and Ni-B systems. The morphological transition which is conditioned by change of a diffusion growth mode on thermal growth (dendrites have the form as a needle) at some undercooling on a dendrite tip ΔT ≥ ΔT* is detected. The *V* versus Δ*T* curve has an S-like character as well as was shown for flat front of crystallization [17]. Values of a critical undercooling ΔT* and a growth velocity discontinuity depend both on a degree of an anisotropy of a kinetic coefficient, and on a difference in energies of activation for atomic kinetics and for diffusion at an interface. The obtained values of a critical undercooling ΔT* and a growth velocity discontinuity on an order of magnitude agree with well-known experimental data [6]. As it has been noted in Ref. [11], that for enough large rates of cooling the transition to a thermal mode can not be realized, i.e. the system "is frozen": growth of crystal is slowed, and melt becomes amorphous (glass transition temperature for the Fe-B system is $T_G$ ~ 0.5 $T_M$ or undercooling is $\Delta T_G$ ~ 1.5 *Q/C* [18]).

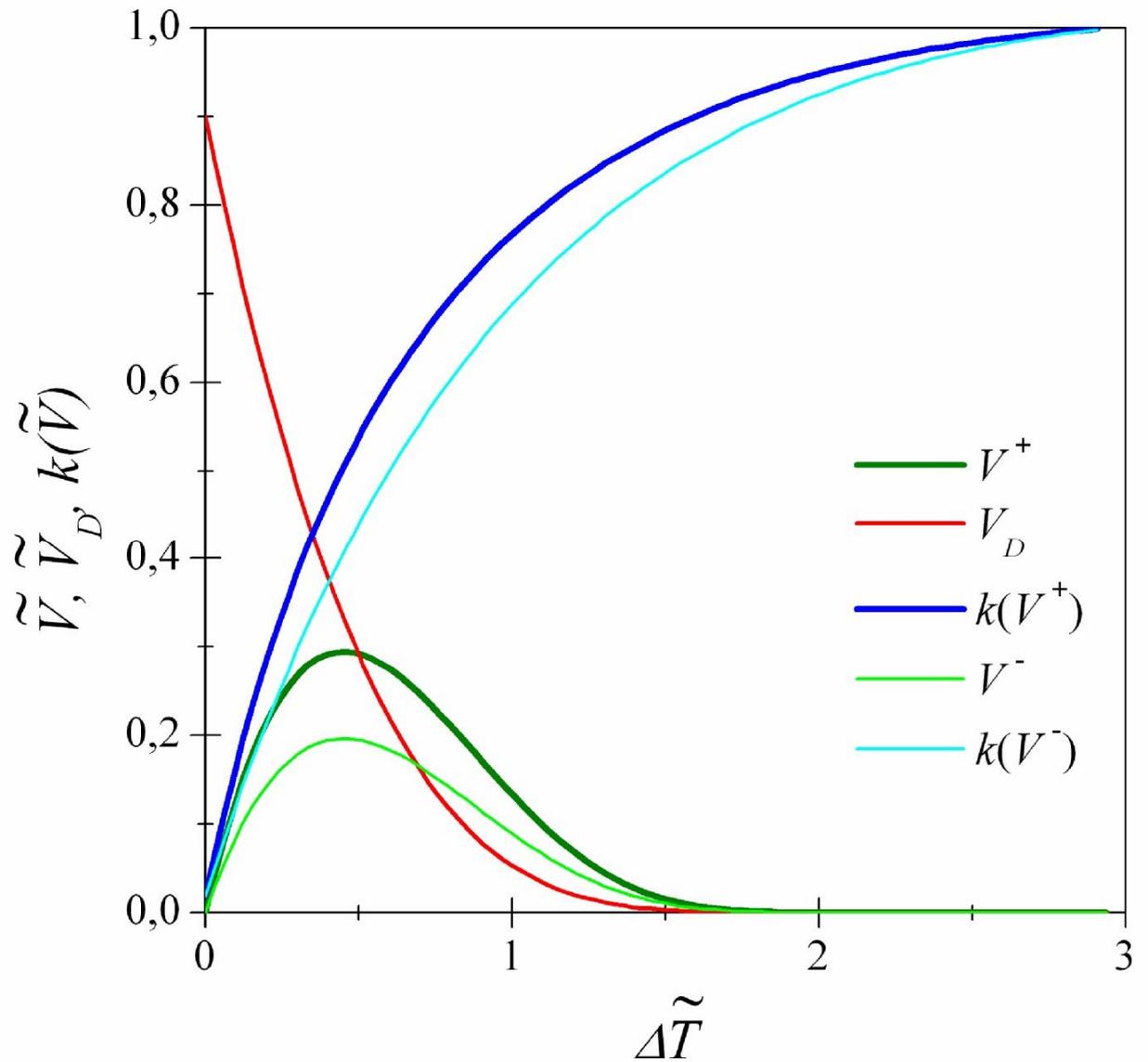

Figure 1: Dependences of velocity of crystal growth $V$, diffusion rate $V_D$ and nonequilibrium partition coefficient $k(V)$ on undercooling $\Delta T$ of the Fe – B melt with $\theta = 2.96$, $E_a/RT_E = 5.55$, $Q/RT_E = 1.0$, $k_e = 0.015$ and $V_D(T_E)/v_0 = 0.9$. Curves was drawn in maximum ($V^+$) and minimum ($V^-$) of orientation dependence of kinetic coefficient (at $V^+/V^- = 1.5$).

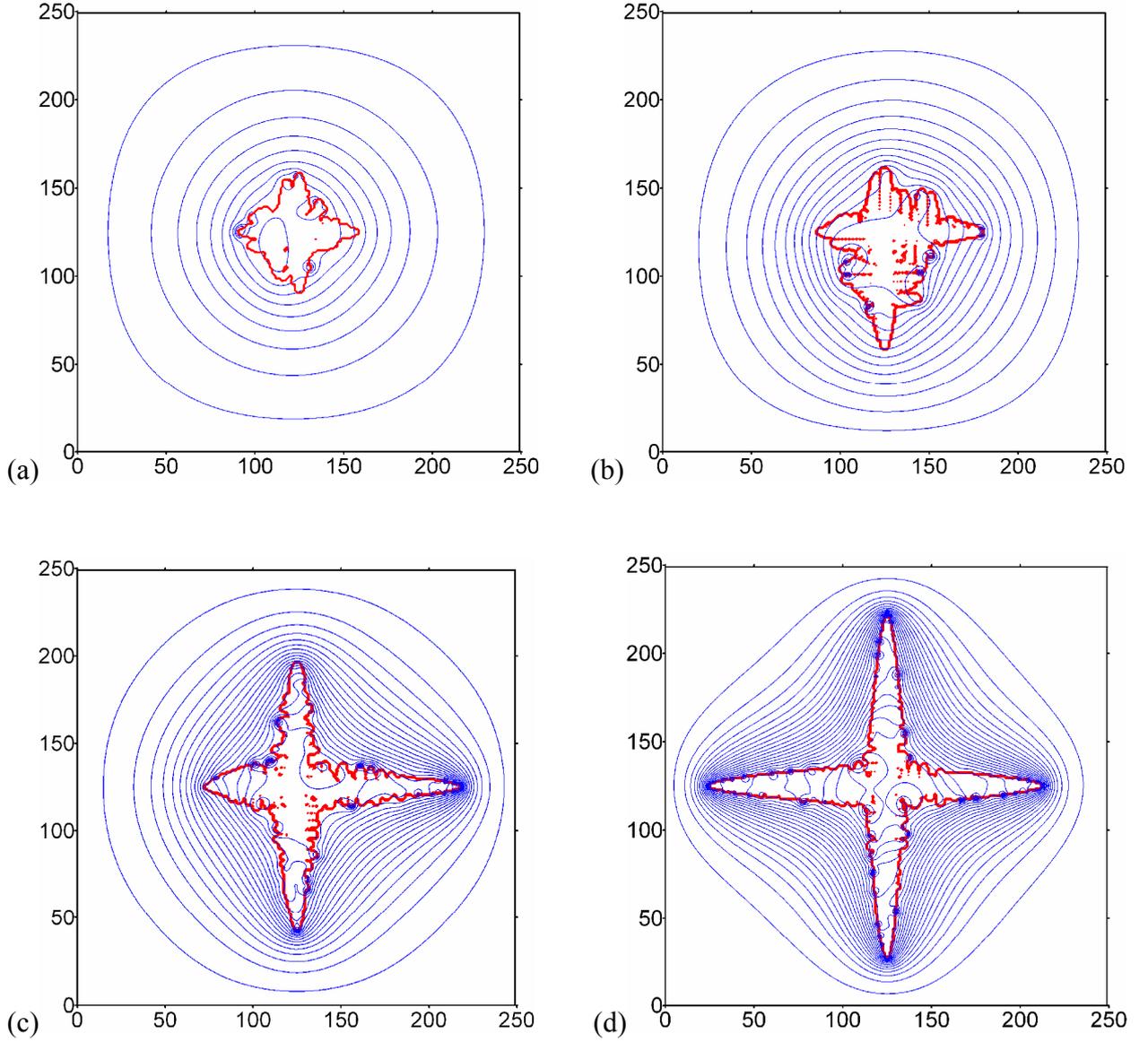

Figure 2: Morphology of dissipative structures (red lines) are formed during Fe-B melt crystallization and temperature fields (blue lines) in the system in some moments of the time $t$ for various values of bath undercooling $\Delta T_{bath}$: (a) $\Delta T_{bath} = 0.55$, $t = 500$; (b) $\Delta T_{bath} = 0.70$, $t = 500$; (c) $\Delta T_{bath} = 0.75$, $t = 250$; (d) $\Delta T_{bath} = 0.85$, $t = 175$.

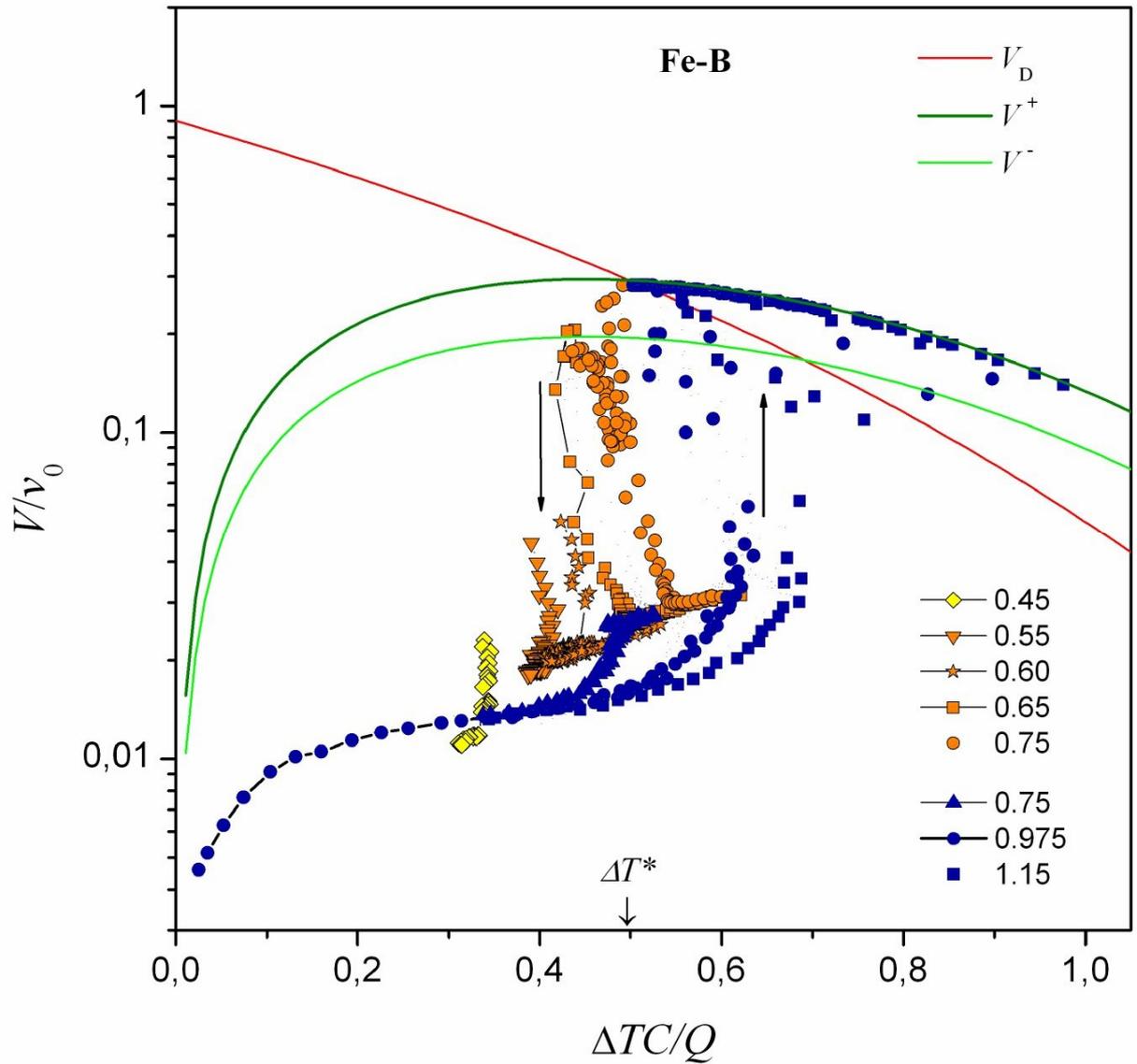

Figure 3: Dependences of the growth velocity on undercooling of the Fe-B melt at dendrite tip. $\Delta T^* = 303$ K. Solid lines: velocity of crystallization in kinetic regime. Blue symbols: results of simulation obtained under melt cooling on the system boundary with the rate R = 0.001 until $\Delta T = \Delta T_B$. Brown symbols: data of computer simulation of system with initial undercooling $\Delta T = \Delta T_{bath}$. Arrows show direction of trajectories of points (in space of variables: $V$, $\Delta T$) during dendrite growth.

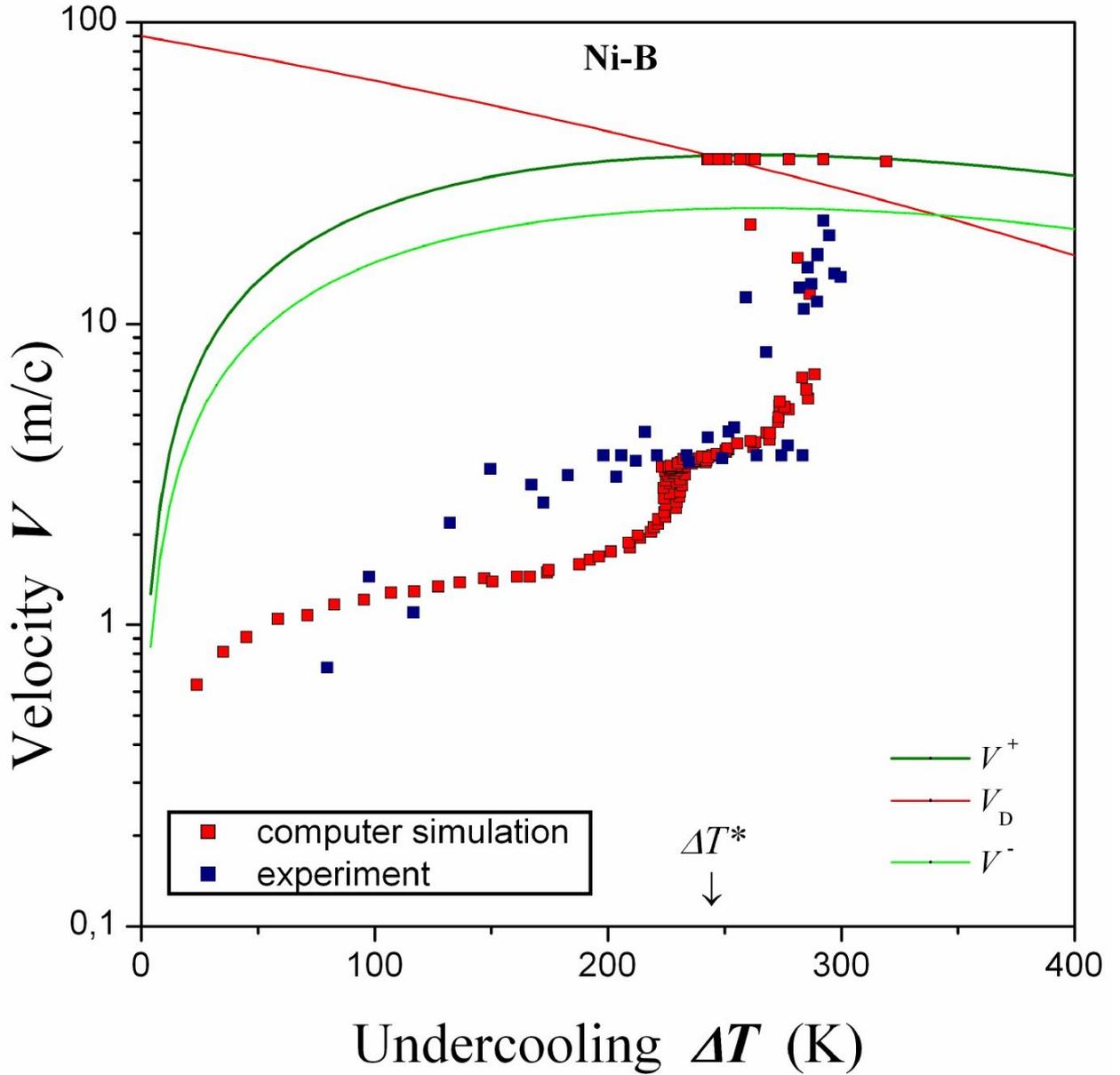

Figure 4: Dependences of the growth velocity on undercooling of the Ni-B melt at dendrite tip. $\Delta T^* = 244$ K. Red symbols: results of simulation obtained under melt cooling on the system boundary with the rate R = 0.001 until $\Delta T_B = 0.9$ with $\theta = 3.651$, $E_a/RT_E = 5.55$, $Q/RT_E = 1.2$, $k_e = 0.015$ and $V_D(T_E)/v_0 = 0.9$; blue symbols: data of experiment for Ni - B.